\documentclass[sor,
]{revtex4-2}

\usepackage{graphicx}
\usepackage{dcolumn}
\usepackage{bm}
\usepackage{physics}

\begin{document}

\preprint{}

\title[]{Single-photon steering}

\author{L M {Ar\'evalo Aguilar}}\email{larevalo@fcfm.buap.mx}\affiliation{Facultad de Ciencias F\'isico Matem\'aticas, Benem\'erita Universidad Aut\'onoma de Puebla.  18 Sur y Avenida San Claudio, Col. San Manuel, C.P: 72520, Puebla, Pue., Mexico}

\date{\today}

\begin{abstract}
According to Schr\"odinger, the laws of quantum mechanics \textit{obliges us to admit that by suitable measurement taken on one of the two system only}\cite{schrodinger} the state of the other system can not only be determined but steered too. That is, it conveys the potential ability to steer the state of another physical system without interacting with it by implementing independent measurements, this nonlocal phenomenon was named \textit{steering}. On the other hand, the first proposals about the nonlocality of a single photon focus on showing the Bell nonlocality by using a single photon path entanglement. This path entanglement of a single photon was also used for analyzing and experimentally produce the steering of a single photon. However, these established facts have been recently called into question suggesting that single-photon entanglement is not non-local.  In this letter, we show that by incorporating and manipulating the internal degrees of freedom of the photon, together with the external path, it is easy to demonstrate the nonlocal effect of steering of a single photon's state. In this sense, the experimental set-up that we propose differs from the one reported in the quantum optics literature which only uses, to the best of our knowledge, the path entanglement of photons for showing the steering phenomenon, i.e. here we exploit the entanglement between the internal and external degrees of freedom of the photon to show this nonlocal effect. The introduction of the photon's internal degrees of freedom in the experimental set-up gives us new insight, advantages and possibilities to control the nonlocal character of the single-photon entangled state.
\end{abstract}

\maketitle

\section{\label{sec:level1}Introduction}

The steering of quantum states was pointed out by Schr\"odinger when he was analyzing the \textit{disconcerting fact} that \textit{the particular choice of observation} affects the wavefunction of the other system even if the measurement was made on one system  only\cite{schrodingerb}, the previous situation arises from the Einstein-Podolsky-Rosen analysis \cite{einstein}. This investigation was pursued in a second paper where he established the steering phenomenon as a \textit{more complete} control, \textit{with the indirect measurement}, of a second system distantly situated \cite{schrodinger}. Additionally, in 2007 steering received an operational definition in quantum information theory as the Alice's  task to convince another distrustful party that she could share an entangled state \cite{wiseman07}. Additionally, there it was demonstrated that steering is a quantum correlation that is stronger than non separability but weaker than Bell nonlocality \cite{wiseman07}, it shows that local hidden state models are unable to characterize the conditional states at one party. Besides, it was shown that steering together with the uncertainty principle underlies nonlocality \cite{hodorecki18}, and that steering is unveiled by measurements of observables that are not jointly measurable \cite{brunner14a,uola15}. Consequently, nowadays we can say that the nonlocal steering phenomenon is a confirmed fact.

It is worth remembering that the first time when the wave particle duality was proposed was using light, this was given in the Einstein-Planck relations $p=\hbar k$ and $E=\hbar\omega$. The second time was when d'Broglie proposed that such equations should be used also for matter. Similarly, the first time that nonlocal quantum correlations was proposed for a single quantum system was the proposal that 
 single-photon entanglement really possesses nonlocal properties. This was theoretically proved first by Tan, Walls and Collet (TWC) \cite{tan}, see also \cite{hardy}, and experimentally proved, free from the detection loophole, by Guerreiro et al. \cite{brunner14b}, where it was clearly demonstrated that single-photon entanglement generates single-photon steering, see also \cite{lee}.  
 
 However, in recent works this possibility has been denied \cite{boyd15,boyd20,Korolkova,Forbes19a,Forbes19b,azzini20,spreeuw}. The perspective employed was conceived using as example the entanglement between the internal degrees of freedom of individual  photons  \cite{boyd15}. Nevertheless, it is not always possible to generalize from a particular photon's state to all entangled states and scenarios. Besides, the nonlocal effects of the single-particle entanglement might be better understood by means of an intelligible  experiment, where single-photon steering could be implemented by using a photon's state that is equivalent to the one analyzed in the single-particle steering case \cite{are20a}, where the steering of a single-particle traversing the Stern-Gerlach experiment (SGE) was proposed. Thereby, the experiment shown below could be thought as an experimental set-up which is easier to implement than the SGE and additionally it could indirectly corroborate the steering in the SGE because the entangled states are strongly similar.  To attain this, the spreading (over large distances) of the entangled wavefunction of the single-photon is employed. Particularly, it will be clear to notice that the nonlocality of the single-photon entangled state allows the photon to "know" the kind of detector that Alice is using to steer Bob's state.

In short, some authors claim that the single-particle entanglement lacks nonlocality \cite{boyd15}: \textit{single-particle entanglement refers to correlations of two different degrees of freedom of an individual particle; this situation cannot lead to nonlocal correlations},  see also \cite{boyd20}. A similar assertion has been made by other different authors \cite{Korolkova,Forbes19a,Forbes19b,azzini20}  taking as main sources the references  \cite{boyd15} and \cite{spreeuw}. For example, in a recent paper the authors claim that entangled states between \textit{two different degrees of freedom of one and the same object} are not  \textit{non-local} \cite{Korolkova} (these authors allude to the state \cite{Korolkova} $\ket{\psi}=(\ket{x_1}\ket{\uparrow}+\ket{x_2}\ket{\downarrow})/\sqrt{2}$ experimentally created by Wineland et al. \cite{wineland}, this state, \textit{is hence prototypic for ‘local entanglement’ of states which cannot be separated spatially}.); this point of view is shared by highly influential journals stating that single-particle entanglement is local only  \cite{Forbes19a,Forbes19b,azzini20}, i.e. lacking nonlocal effects; additionally see a recent paper that also endorses the claim that single-particle entanglement just serves to text noncontextual reality models \cite{wang21}. 

Hence, the previous paragraphs strongly suggest that nonlocal correlations can only occur in entangled states comprising  two quantum systems. Additionally, it may result counter-intuitive to name a phenomenon \textit{steering}, which implies nonlocal correlations, when it comes from a single-particle. However, the fact is that this effect was theoretically proved \cite{are20a} and in the case of single-photon steering it was experimentally proved also\cite{brunner14b}.

By restricting oneself  to the particular kind of entangled states considered in reference \cite{boyd15}, i.e. bipartite entanglement between the polarization and the orbital angular momentum (OAM) of photons,  it could lead us to conclude that single-particle entanglement is not nonlocal. However, this assertion based on a particular kind of bipartite entangled states could not be generalized to all entangled states or all scenarios. More important, if we remain embracing the idea that single-particle entangled states do not possess non-local correlations, then many of the non-local task such as quantum key distribution might not be investigated or even implemented by using these resources.

In recent years world leading quantum groups have made a great effort to experimentally demonstrate single-photon steering \cite{are20a}. For example, Fuwa et al. demonstrated the violation of a steering inequality employing a single-photon entangled state by implementing homodyne detection \cite{fuwa15}, additionally see also \cite{lee,garrisi,guerreiro12,george13}; and Guerreiro et al. have proven the violation of a new steering inequality (employing a single-photon entangled state) free of detection loopholes using as the set of assemblage the eigenstates of the displaced operator \cite{brunner14b}.  Besides, recently it was theoretically shown that single-particle entanglement between an internal DoF and an external DoF produces steering in atoms crossing the famous Stern-Gerlach experiment \cite{are20a}. These experiments \cite{fuwa15,garrisi,guerreiro12,george13,brunner14b,lee} together with the theoretical proposals given in references \cite{are20a,are20b} cast doubts on whether the assertion that \textit{single-particle entanglement cannot produce nonlocal quantum correlations} is a valid generalization.

In this letter, we show that the nonlocal effect of single-photon steering could be produced by using an alternative experimental  scheme with a state that is similar to the one that was proposed in the single-particle steering case \cite{are20a}; that is to say, the entanglement between an internal degree of freedom of the photon and its external path. In particular, from this scheme it is easy to see that the nonlocality of the single-photon entangled state allows  Alice to steer Bob's state depending on the kind of measurement apparatuses that she employs. The states that we use in this letter differ from the state used in the first proposal of single photon nonlocality by TWC \cite{tan}  who use as an initial state $\ket{1}_u\ket{0}_v$, which is transformed after the first  beam splitter as: $\ket{1}_{b_1}\ket{0}_{b_2}+\ket{0}_{b_1}\ket{1}_{b_2}$; and the state used by Hardy \cite{hardy} who uses as an initial state $(\ket{1}_s+\ket{0})_s\ket{0}_t$, which is transformed after a first  beam splitter as: $q\ket{0}_{u_1}\ket{0}_{u_2}+\frac{ir}{\sqrt{2}}\ket{1}_{u_1}\ket{0}_{u_2}+\frac{r}{\sqrt{2}}\ket{0}_{u_1}\ket{1}_{u_2}$, due to the fact that they use the entanglement in the photon path only. In this work, alternatively, instead of using the single-photon path entanglement only, we devise a proposal which take into account the internal degree of freedom of the photon and we show that the steering of the single-photon is clearly and effortlessly conceived. On the other hand, in recent interesting works by Das et al. \cite{das21a,das21b}, who also employ the entangled path state $\ket{1}_{b_1}\ket{0}_{b_2}+\ket{0}_{b_1}\ket{1}_{b_2}$, it is argued that the experimental set-up (i.e. the measurement parts) used by TWC does not reliable test the violation of the Bell inequality and that a local hidden variable model could reproduce the thought nonlocal events considered by TWC; besides, these authors claim that the initial Hardy's state and his proposed experimental set-up could reliable produce nonlocal effects \cite{das21a}.  Essentially, the works by TWC, Hardy and Das et al. rely on the path entanglement of the single photon (although they differ on the measurement set-up), whereas  the proposal carried out in this letter relies on the entanglement between an internal degree of freedom of the photon with the potential paths that the photon follows. Notice that what Das et al. demonstrated is that the experimental set-up together with the kind of Bell inequality that  was used by TWC do not allow a test of the violation of the CHCH-Bell inequality due to both: i) the assumption that the total intensity given by $I_j(\lambda)=I_{c_j}(\theta_j,\lambda)+I_{d_j}(\theta_j,\lambda)$ is independent of the local value of $\theta_j$, and ii) the use of a constant amplitude of the weak local field used in the homodyne measurement, see \cite{das21a} for details. Besides, it is worth of mention that the works carried out in references \cite{brunner14b,fuwa15} are variants of the TWC analysis.

On the other hand, in references \cite{elena2018,are20a} it was proved that the entanglement between the internal and external degrees of freedom of a particle violates the CHSH-Bell inequality; hence, as the state that we employ in this letter is a similar entangled state between the internal and external degrees of freedom it would also violate such inequality. Hence, we will exploit the entanglement between the internal and external degrees of freedom to show the steering of a single photon state; this approach, to the best of our knowledge, has not been previously proposed jet.


\section{The scheme}

\begin{figure}[h!]
   \centering
  \includegraphics[width=120mm]{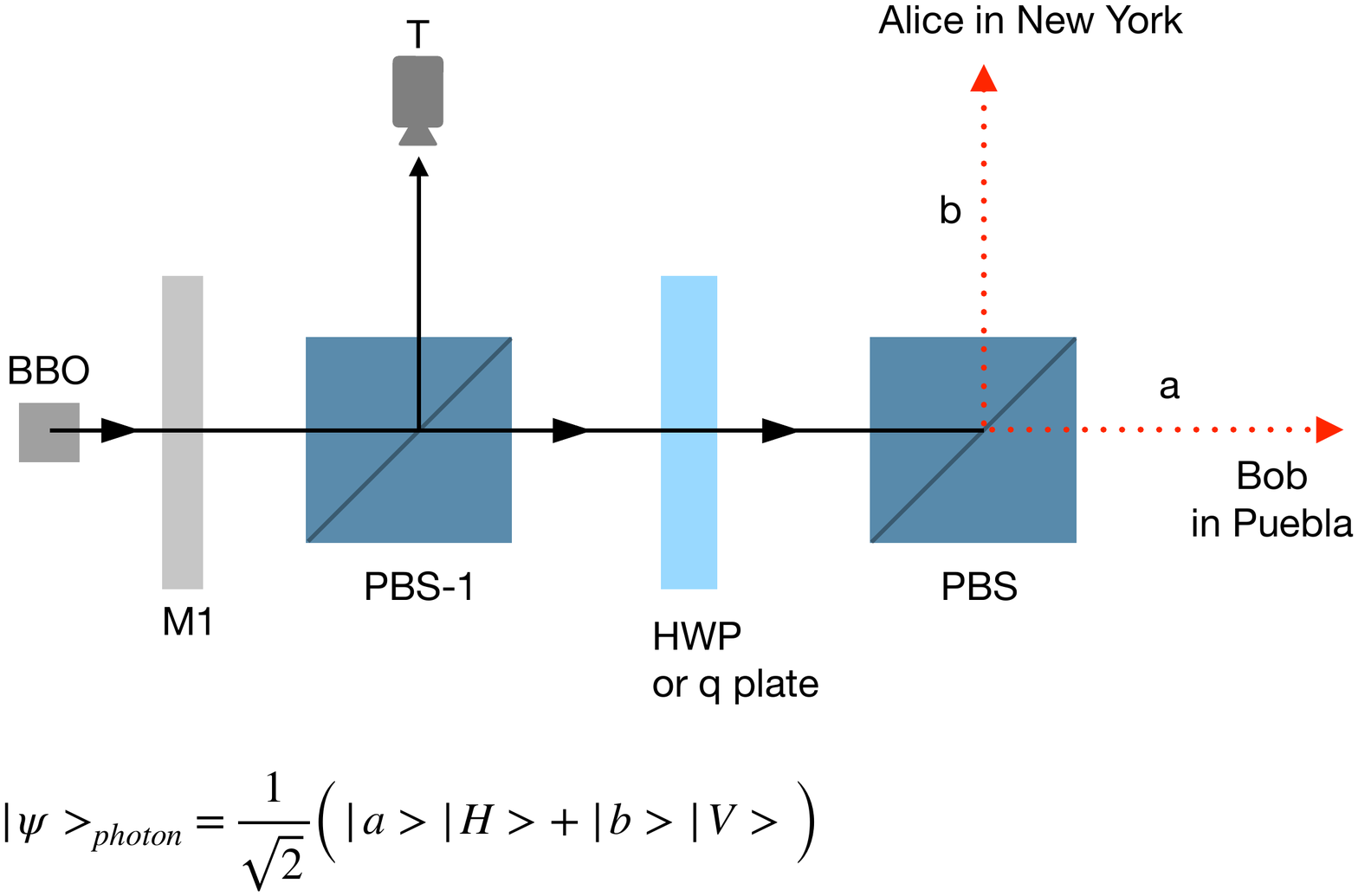}
\caption{\label{puebla} A single-photon steering experiment, where Alice is able to steer Bob's state at a spacelike distance between the cities of New York and Puebla.}
\end{figure}
The scheme that is used to help the explanation of the single-photon steering is shown in Fig. (\ref{puebla}). It takes as base the schemes given in references \cite{kim03,are20a} and it has as a main goal to prepare the state given by the following equation:

\begin{equation}
\ket{\psi}_{photon}=\frac{1}{\sqrt{2}}\left(\ket{a}\ket{H} +\ket{b}\ket{V}\right),
\label{sp-entangled}
\end{equation}
where $\ket{a}$ and $\ket{b}$ are the probable paths that the photon could follow, towards the cities of New York or Puebla; $\ket{H}$ and $\ket{V}$ are the vertical and horizontal polarizations of the photon, respectively. Usually, in quantum optics $\ket{a}=\ket{0}\ket{1}$ and  $\ket{b}=\ket{1}\ket{0}$.  Eq. (\ref{sp-entangled}) represents a single-photon entangled state where the dichotomic entangled variables are the orthogonal polarizations $\ket{H}$ and $\ket{V}$  and the spatial modes $\ket{a}$ and $\ket{b}$ --as the phase produced by the reflection in the Beam splitter could be compensated by a linear optical component we do not write it down--. The preparation of the single-photon entangled state given by Eq. (\ref{sp-entangled}) is as follows: First, a single photon is generated by using a postselection method where a type-II BBO crystal is pumped creating  two orthogonal polarized photons propagating in the same direction, then a Polarized Beam Splitter, PBS-1 in the Fig. (\ref{puebla}), divert the vertical polarized photon towards the trigger detector T. Hence, the trigger signal indicates that a single photon with horizontal polarization was created. Second, the half-wave plate (HWP) rotates the polarization of the horizontal polarized photon to $45^{\circ}$ just before the second PBS, then the latter beam splitter (PBS) produces the entangled state given by Eq. (\ref{sp-entangled}).

The state given by Eq. (\ref{sp-entangled}) is an entangled state between the internal (polarization) and external (path) DoFs of the photon, it is equivalent to the following entangled state given in references \cite{are2017a,are2017b}:
\begin{eqnarray}
\label{psifinal}
\ket{\psi(t)} =C_0M(x,y)\Big(\ket{\psi_+}\ket{\uparrow_z}+\ket{\psi_-}\ket{\downarrow_z}\Big),
\end{eqnarray}
where
\begin{equation}
\braket{z}{\psi_{\pm}}=
		e^{\frac{\mp it\mu_c}{\hbar}(B_{0}+bz)}
		\exp\left[\frac{-1}{4(\sigma_{0}^{2}+i t\hbar/2m)}\left(z\pm \frac{t^{2}\mu_c b}{2m}\right)^{2}\right],
\end{equation}
\begin{equation}
C_0=e^{\frac{-it^3\mu_c^2b^2}{6m\hbar}}\frac{1}{\sqrt{2}}
		\left[\frac{\sigma_{0}}{(2\pi)^{1/2}}\right]^{3/2}
		\left(\sigma_{0}^{2}+\frac{i \hbar t}{2m}\right)^{-3/2},
\end{equation}
and
$M(x,y)=e^{-\sigma_0^2k_y^2}
	 e^{\frac{ 4y\sigma_{0}^{2}k_{y}}{4(\sigma_{0}^{2}+ t\hbar/2m)}}
\exp\left[\frac{-(x^{2}+y^{2}-4\sigma_{0}^{4}k_{y}^{2})}{4(\sigma_{0}^{2}+ it\hbar/2m)}\right].$
This state is generated in the famous Stern-Gerlach experiment, and it represents an entangled state between the internal and external DoFs of a single particle.  It is worthy of mention that it was already demonstrated that the entangled state given by 
this equation violates the  Bell's Clauser–Horne–Shimony–Holt (CHSH-Bell) kind inequality \cite{elena2018}.

The entangled state of the single-photon given by Eq. (\ref{sp-entangled}) resembles the Einstein's boxes with entangled states given in reference \cite{are20a}, an it allows Alice to steer Bob's sates. To perceive this, notice that if Alice decided to measure the polarization of her system then the following two possibilities would occur:

\begin{enumerate}
\item If Alice got as result that her photon has a vertical polarization, the wave function would collapse towards $\ket{b}\ket{V}$, with the photon at New York city. That is, as $\ket{b}=\ket{1}\ket{0}$, Alice gets the $\ket{1}$ state and Bob gets the $\ket{0}$ state.
\item However, if Alice detected noting, the state vector would collapse towards $\ket{a}\ket{H}$ with the photon at Puebla city. That is, as $\ket{a}=\ket{0}\ket{1}$, Alice gets the $\ket{0}$ state and Bob gets the $\ket{1}$ state.
\end{enumerate}

The effect of "detect nothing" could be explained in terms of the Einstein's boxes, where a party could collapse the state of another party by open a box and finding nothing, see reference \cite{are20a}. Also, notice that in the previous consideration Alice is measuring the polarization of the photon; however, if she measured the presence of the photon \cite{lee} instead she would obtain similar states, but at the cost of destroying the photon.

On the other hand, if Alice decided to measure in different bases, for example in the bases $\frac{1}{\sqrt{2}}(\ket{V}\pm\ket{H})$ or in the left and right circular polarization, in the former case the following possibilities arise:
\begin{enumerate}
\item If Alice got the polarization $\ket{V}+\ket{H}$, the spatial part of the state wavefunction would collapse towards $\ket{a}+\ket{b}$, i.e. a superposition of "being" in New York and Puebla at the same time. Remember that  $\ket{a}+\ket{b}=\ket{0}\ket{1}+\ket{1}\ket{0}$; that is to say, what Bob gets is an  entangled photon state, i.e. the local Bob's quantum state is $\hat{\rho}_B=Tr_A\{\hat{\rho}_{AB}\}=\ket{0}\bra{0}+\ket{1}\bra{1}$.
\item If Alice got the polarization $\ket{V}-\ket{H}$, the spatial part of the  wavefunction would collapse towards $\ket{a}-\ket{b}$, i.e. a superposition of "being" in New York and Puebla at the same time but with a different relative phase than the previous case. Please take into account that  $\ket{a}-\ket{b}=\ket{0}\ket{1}-\ket{1}\ket{0}$; that is to say, what Bob gets is an  entangled photon state, i.e. the local Bob's quantum state is $\hat{\rho}_B=Tr_A\{\hat{\rho}_{AB}\}=\ket{0}\bra{0}-\ket{1}\bra{1}$.
\end{enumerate}
similar results (but with different phases) would be achieved if Alice decide to measure in the left and right circular polarization bases.

Therefore, by choosing the measurement  bases Alice is allowed to steer Bob's state at two space-like separate places like New York city and Puebla city. As it is well known, steering a quantum state means that a nonlocal resource is being used, since steering is a nonlocal property that lies between Bell nonlocality and entanglement \cite{wiseman07,jones07,cavalcanti09,uola20,brunner14,cao}. It is worthy of mention that the measurements could be made at two different places, i.e. New York city or Puebla city. Hence, if Bob also make measurements, then Bob and Alice would obtain similar or different results   and process depending on whether or not they use generalized or projective measurements \cite{arno}. Therefore, in the single-photon entangled state that is customized adapted in the above scheme, the non-local features arises from independent measurements at two faraway places.

Additionally, is worth of noting that if we replaced the HWP of Fig. (\ref{puebla}) by a q plate \cite{Nagali,Marrucci}, the q plate would produce a single-particle entangled state between the left and right circular polarization and the orbital angular momentum, $\ket{\psi}=\frac{1}{\sqrt{2}}(\ket{L}_\pi\ket{-2}_l+\ket{R}_\pi\ket{2}_l) $ \cite{Nagali}, of the kind mentioned in reference \cite{boyd15}, where $\ket{R}_\pi$ and $\ket{L}_\pi$ are the right and left circular polarization  $\ket{2}_l$ and $\ket{-2}_l$ are the angular momentum sates. Hence, by writing this state in terms of the horizontal and vertical polarization we can ascertain the kind of state that is produced after the photon traverse the q plate. Then, the state before the second PBS is $\ket{\psi}=\frac{1}{2}\Big[\ket{H}_\pi\big(\ket{+2}_l+\ket{-2}_l\big)+ i\ket{V}_\pi\big(\ket{+2}_l-\ket{-2}_l\big)\Big]$

Therefore, after traversing the second PBS the entangled state evolves towards:  
$\ket{\psi}=\frac{1}{2}\Big[\ket{a}\ket{H}_\pi\big(\ket{+2}_l+\ket{-2}_l\big)+ i\ket{b}\ket{V}_\pi\big(\ket{+2}_l-\ket{-2}_l\big)\Big]. \label{three}$
Thus, this produces a similar situation that the one reported above which allowed Alice to steer Bob's state by means of suitable measurements. But in this case, it is an entangled state between three DoFs of the photon. Hence, Alice is able to measure in any bases of the three sub-spaces: i) the path, ii) the polarization, or iii) the orbital angular momentum (OAM); this scenario gives her more advantages than in the bipartite case. Notice that two DoFs of this state are the internal polarization and the OAM, hence by measuring in any of these two bases Alice may steer Bob's state. Of course, if Alice used a bipartite entangled state comprising the polarization and the OAM only, she would not be able to steer Bob's state, as the authors of reference \cite{boyd15} rightly  imply. One of the drawbacks of reference \cite{boyd15} is perhaps that the authors generalize this single situation for disproving the existence of single-photon nonlocal effects. However, as it is well known it is not always valid to generalize from a single setting, especially when one try to prove nonlocality with two internal degrees of freedom; on the other hand, by means of the scheme given above we could easily explain the phenomena of single-photon steering. 

We can draw some conclusions from our discussion: as it is well known, quantum correlations are significantly different from the classical ones; hence, we agree that classical states cannot generate true quantum entanglement, thus the  use of the name \textit{classical entanglement} for alluding to classical non-separable states could lead to confusion. On the other hand, quantum single-particle entanglement could legitimately address many issues of quantum theory regarding the applications and understanding of fundamental questions on nonlocal quantum correlations and quantum mechanics. In this letter, we have proposed a different experimental set-up to show nonlocal effects in a single photon. In this new alternative set-up, it was shown that by manipulating the internal and external degrees of freedom of the single photon, the steering effect could be achieved. Particularly, the nonlocal features of the single-photon was addressed by using  both i) a single internal and a single external degrees of freedom and ii) two internal degrees of freedom together with an external degree of freedom. These configurations differ from the ones reported in literature to show the nonlocality of the photon which  use,  as far as we know, the path entanglement of photons only. Hence, if one take into account the conclusions reached in references \cite{das21a,das21b} (that is, that the correlations measured by the TWC's experimental set-up are explained by a local hidden variable model), then the experimental set-up proposed here, could be a good alternative to demonstrate the nonlocality of a single photon. Additionally, it is worth mentioning that this scenario differs from the set-up shown in reference \cite{are20b} where two external degree of freedom are entangled with an internal degree of freedom. These two kinds of entanglement represent two different experimental situations which produce different physical results; for example, the possibility of considering two additional observers in the case of reference \cite{are20b} (i.e two external degrees of freedom entangled with an internal degree of freedom) is not possible in the photon case used in this letter where two internal degrees of freedom are entangled with an external degree of freedom. Finally, an additional difference is that in the case of the photon, the vacuum state $\ket{0}$ plays a significant role (for example it could allow quantum tomography ) that is not present in the situation that was addressed in reference \cite{are20a}. In conclusion, the nonlocal steering phenomenon could be facilitated by allowing entanglement between internal and external degrees of freedom in a single photon.




\begin{acknowledgments}
We thank Alba Julita Chiyopa Robledo for the proffreading of the manuscript. There was not any support of any scientific agency. We thank Vicerrectoría de Investigación y Estudios de Postgrado from BUAP for providing us with computer equipment to carry out our academic and research duties.
\end{acknowledgments}

\bibliography{plain}

\end{document}